# MODELING MUSIC MODALITY WITH A KEY-CLASS INVARIANT PITCH CHROMA CNN

**Anders Elowsson**
KTH Royal Institute of Technology
anderselowsson@gmail.com

**Anders Friberg**
KTH Royal Institute of Technology
afriberg@kth.se

## ABSTRACT

This paper presents a convolutional neural network (CNN) that uses input from a polyphonic pitch estimation system to predict perceived minor/major modality in music audio. The pitch activation input is structured to allow the first CNN layer to compute two pitch chromas focused on different octaves. The following layers perform harmony analysis across chroma and time scales. Through max pooling across pitch, the CNN becomes invariant with regards to the key class (i.e., key disregarding mode) of the music. A multilayer perceptron combines the modality activation output with spectral features for the final prediction. The study uses a dataset of 203 excerpts rated by around 20 listeners each, a small challenging data size requiring a carefully designed parameter sharing. With an $R^2$ of about 0.71, the system clearly outperforms previous systems as well as individual human listeners. A final ablation study highlights the importance of using pitch activations processed across longer time scales, and using pooling to facilitate invariance with regards to the key class.

## 1. INTRODUCTION

### 1.1 Modality

Minor and major modality is a function of scale, harmony and tonality and is perceptible even to very young children [20]. However, the rich variability of music harmony renders many compositions hard to classify into a minor or major mode. Researchers have therefore investigated modality as a continuous variable in listening tests, producing more or less uniformly distributed averages with high internal consistency. Such a continuous variable, ranging from minor to major, has interchangeably been referred to as *modality* [12-14, 28, 34], *mode* [1], *key mode* [33], *mode majorness* [33], and *majorness* [2, 28, 33]. We will mainly use the term "modality" or "minor/major modality".[1] This paper aims to improve on previous methodologies for predicting perceived modality, designing a CNN that is able to model associated intricacies of musical harmony.

In a listener study [14], rated modality had a significant correlation (0.3-0.6) with rated *speed*, *articulation*, *pitch* (low/high) and *timbre/brightness* – happy tunes in major mode are likely more often performed with a higher articulation (staccato). This means that a system can be designed to predict perceived modality simply by picking up aspects in the audio not directly associated with harmony. Music information retrieval (MIR) systems relying on such confounding factors of variation have been challenged by Sturm [37]. The CNN architecture proposed in this study tries to minimize these interactions by specifically targeting properties directly linked to modality, as expanded upon, e.g., in Section 2.4.

### 1.2 Previous Studies Predicting Modality

Two previous studies have attempted to predict perceived modality from music audio. The first study [14] used partial least squares regression applied to audio features from the MIR toolbox [28, 29]. Two models were tried, the first using dedicated modality features and the second also including other spectral features. They were evaluated on the same two datasets used in the present study (Section 5.1), reaching an $R^2$ of 0.43 (0.38) and 0.47 (0.53) respectively (results for the second model in parenthesis).

A second study [1] have instead used the Inception v3 architecture [38] applied to a mel-frequency spectrogram. Results on a dataset of 5000 15 seconds (s) excerpts with lower ground truth consistency was $R^2 = 0.23$ (based on the Pearson's correlation coefficient of 0.48 reported in an additional/supplemental paper [2]). The model was developed to handle numerous perceptual features and may not be ideal for modality; the pooling operations applied across mel-frequency obfuscates tonal interrelationships at ranges larger than the pooling kernels. Since the filters span the time dimension, the model may to some extent instead make predictions from other aspects of the audio that covaries with modality, as outlined in Section 1.1.

### 1.3 Pitch Chroma and Deep Layered Learning

Chroma features, imposing spectral energies across a wide frequency range onto the twelve pitch classes of a musical octave, have a long tradition in MIR [15, 17, 35]. Pitch chromas have also been derived both from MIDI data and estimated through the autocorrelation function in the past [39]. A problem with the chroma is that it often is affected by interferences and becomes noisy [23]. Researchers have used various techniques to mitigate these issues, including harmonic percussive source separation [40] and cepstral whitening [31]. A multilayer perceptron (MLP) has also been used, with chord annotations defining ground truth pitch classes [23]. In this paper, we instead use the output from a high accuracy/resolution polyphonic pitch tracking system [8]. The pitch transcription is reshaped and fed as input to a CNN so that several pitch chromas emphasizing

---

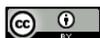

[1] Since "mode" and "modality" have a wider scope, perhaps "modalite" could be a useful nomenclature for future studies.

different octaves can be learned within its first layer. The full architecture thus uses intermediate targets to restructure the learning problem according to the inherent organization of music. Such a "deep layered learning" approach [6], learning intermediate equivariant music representations, has been used for various MIR problems in recent years [e.g., 5, 19, 26].

### 1.4 CNNs and Ensemble Learning

The CNN proposed in this study use filter kernels operating across pitch/pitch class, pitch octaves and time scales, applying the same processing to each time frame. Previous CNNs for harmony processing have instead used filter kernels operating across time and frequency, for tasks such as key estimation [22, 25] and chord recognition [18, 24].

A CNN trained two times with randomly initialized parameters will generally produce two different predictions to the same input data. This is something that makes neural networks useful for ensemble learning [16]. The average (ensemble) prediction from several models containing more or less decorrelated errors will be better than randomly choosing one of them [32, 36]. Ensemble learning is used in this paper, as specified in Section 4.

### 1.5 Overview of the Paper

This paper presents a pitch chroma CNN architecture for predicting perceived modality. Section 2 describes how the pitch activation input to the CNN is computed and structured. In Section 3, the network architecture and training procedure is outlined. Section 4 describes how several CNNs were combined into an ensemble, and how a global prediction was made using additional features in an ensemble of MLPs. The two datasets and the evaluation procedure is described in Section 5, and results presented in Section 6. Section 7 presents an ablation study, testing how the design of the model affect predictive performance, and Section 8 offers conclusions.

## 2. PREPARING THE INPUT REPRESENTATION

### 2.1 Defining a Start and End Time for Each Excerpt

A start and end time were first determined for each musical excerpt (ME) so that the CNN would not have to make any predictions for silent parts in the beginning and end. A magnitude log-frequency spectrogram with 60 bins per octave was computed as described in [8] (pre-filtering). Let $x$ be a vector representing the frequency response in time frame $i$. The overall magnitude of that time frame, across all frequency bins, was then defined as

$$m_i = \sqrt{\overline{x^2}}, \qquad (1)$$

using the elementwise square and arithmetic mean, and forming $m$ as a vector across time. The signal level was defined as $L_i = 20 \log_{10} m_i$, and the resulting vector filtered with a Hann window of width 61 frames (0.35 s). The *average* signal level of the ME was instead defined as $L_a = 20 \log_{10} \overline{m}$. The first time frame with a signal level within 10 dB of $L_a$ defined the start, and the last frame within 10 dB of $L_a$ defined the end of the ME.

### 2.2 Pitchogram

The input to the CNN was extracted from the *Pitchogram* representation computed with an existing machine learning system [8]. That system uses two stages to compute the Pitchogram. First, a sparse filter kernel operates across a log-frequency spectrogram to compute activations corresponding to tentative fundamental frequencies ($f_0$s), up-sampled through parabolic interpolation to a centitone resolution. These tentative $f_0$s are then analyzed in a deeper network and computed activations inserted at the corresponding pitch bin in the Pitchogram. The Pitchogram thus contains $f_0$ activations and has a pitch resolution of 1 cent/bin and a time resolution of 5.8 ms/frame.

### 2.3 Extracting Semitone-spaced Pitch Vectors

The Pitchogram was down-sampled to 1 bin/semitone before processing by the modality CNN. To do this, a Hann window of height 141 bins (cents) was first applied across pitch to smooth the pitch response. Then, to adjust MEs that deviate globally from standard tuning (specifying A4 to a frequency of 440 Hz), the Pitchogram of each ME was "tuned". This was especially important for some of the MEs from the film music datasets (Section 5.1), where the orchestral performances had different tunings. The tuning was achieved by locating the maximum activation in a vector $v$ of length 100, where each element corresponds to the sum of pitchogram activations at a specific cent value (i.e., all bins 47 cents above standard tuning were summed as the $47^{th}$ entry of the vector). Only one vector was computed for each ME (i.e., global tuning). The whole Pitchogram was then shifted $\pm 50$ cent based on the index of the maximum element. Finally, semitone-spaced activations were extracted between MIDI pitch 26-96, resulting in a pitch vector for each frame of height 71.

### 2.4 Varying Time Scales

We assume that listeners use pitch information at varying time scales to form an overall impression of the modality of a piece of music. At the shortest time scale, concurrent tones can form harmonic relationships that sound more like a major chord or a minor chord. At slightly longer time scales, tones played in succession may together imply the mode of the chord. At even longer time scales, the combination of tones and their relative activation may resemble key profiles/tonal hierarchies [27] that are more or less indicative of a major or minor tone scale. We wanted to develop a model that was agnostic to various factors that may covary with modality, such as accentuation (see Section 1.1), in the hope that our model would then generalize better to other datasets lacking these covariations. This precluded many models tracking variations in pitch activations across time (e.g., recurrent neural networks). Therefore, in order to still account for tonal relationships evolving over time, the pitch response was smoothed across time with filters of varying width, producing pitch vectors responding at varying time scales. The smoothing was done with Hann windows of width

$$w = 10 \times 3^n + 1 \qquad (2)$$

where *n* varied between 1-5. The shortest Hann window therefore had a width of 31 frames (0.18 s) and the longest a width of 2431 frames (14.1 s). Since the unfiltered pitch vector was also included, the processing was applied at six different time scales. The smoothed pitch vectors were finally stacked across width, as shown in the left pane of Figure 1. This enabled the system to combine them during processing with a filter of width and stride 6 (Section 3.1).

### 2.5 Octave Spaced Depth (Chroma)

The proposed CNN computes a pitch chroma within its first layer. To facilitate this, the pitch vector was divided into 5 overlapping sections spaced an octave apart, each covering 23 semitones. The sections were concatenated across depth, resulting in aligned pitch classes (facilitating chroma processing within the CNN with filters extending across depth). This depth dimension is illustrated for a single time frame in the right pane of Figure 1. The CNN input of each time frame was thus prepared. It can be understood as a $23 \times 6 \times 5$ three-dimensional tensor, where height (23) represents pitch class, width (6) represents different time scales and depth (5) represents pitch octaves.

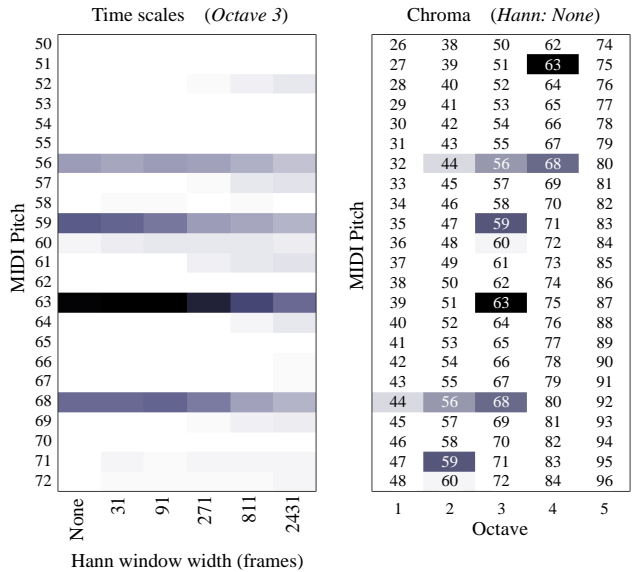

**Figure 1.** Two dimensions of the 3D-tensor input to the CNN for a single time frame, consisting of pitch activation from [8] restructured across pitch, time scale and pitch class. The time frame is taken 8 seconds into ME No 4 in the film clips dataset, where a G♯ minor chord was played. The left pane consists of activations at varying time scales in the third octave. The right pane consists of the different pitch octaves, shown at the shortest time scale.

### 2.6 Segmentation

Each ME was divided into 6 overlapping segments of length 9 seconds (1550 time frames). Since no ME was longer than 54 seconds, these segments spanned the entire ME. Each segment of an ME was assigned the same ground truth annotation that had been established from listener ratings (Section 5.1). Since each frame (input tensor) had $23 \times 6 \times 5 = 690$ input values, each segment had slightly above a million input values ($690 \times 1550$).

## 3. CNN ARCHITECTURE AND TRAINING

### 3.1 CNN Architecture

The CNN that was applied to each 9 seconds segment is shown in Figure 2. The same processing architecture was applied to each time frame of the segment. Convolutional filters always operated across unspanned dimensions, and zero-padding was never utilized, thereby shrinking the output space when applying the filters. Rectified linear units (ReLUs) were used as activation functions. Since the dataset in the study was small (203 MEs), it was important (and a challenge) to keep the number of learnable parameters small while retaining the ability to model the intricacies of musical harmony. The network had a total of 413 learnable parameters, including parameters for the batch-normalization that was applied after each ReLU layer.

As shown in the figure, the first *chroma layer* is used for converting input tensors to pitch chromas. Two filters learn weights for each octave, operating across pitch (height) and time scales (width). The two resulting pitch chromas are split into two branches, and processed with 1 and 5 filters respectively in the subsequent *harmony analysis* layer. This was done to reduce the total number of parameters while still achieving the following objectives:

- Allow the system to use two pitch chromas so that it could, for example, independently filter and account for information in the bass and higher pitches.
- Output 6 different activations for each time scale and pitch class from the *harmony analysis* layer using only $12 \times 6 + 6 = 78$ filter parameters.

Each filter used for harmony analysis spans an octave so that all pitch classes are taken into account when computing an activation (attempts at dividing the harmony analysis into two layers with shorter filters gave slightly lower results). Since the input has a height of 23, each filter is applied at 12 positions when operating across pitch, one for each pitch class. Thus, the input and filter sizes are the minimum sizes for which the same combination of all pitch classes at various keys can be subjected to the same filters. The two branches were concatenated across depth and the next filter (orange) then spanned the various *time scales*. The subsequent *max pooling* filter (purple) provides invariance with regards to the key class that the music was performed in, since it spans all 12 pitch/key classes. Based on the max pooling design, it can be expected that the previous layers of the CNN learn to either react to minor or major chords, minor or major tonal hierarchies, or various combinations thereof. The strongest indications of both major and minor modality across key are then passed to the subsequent *fully connected* (locally) layer (orange) for further processing. This layer is implemented as 7 filters that span the entire local input space (a fully connected layer would instead span the entire segment, something that is not desirable). A second filter (orange) combines the previous activations into a single *frame prediction*.

Finally, *average pooling* (purple) of the frame predictions is applied across the entire 9 seconds segment to produce a *segment prediction* (red) as a regression output.

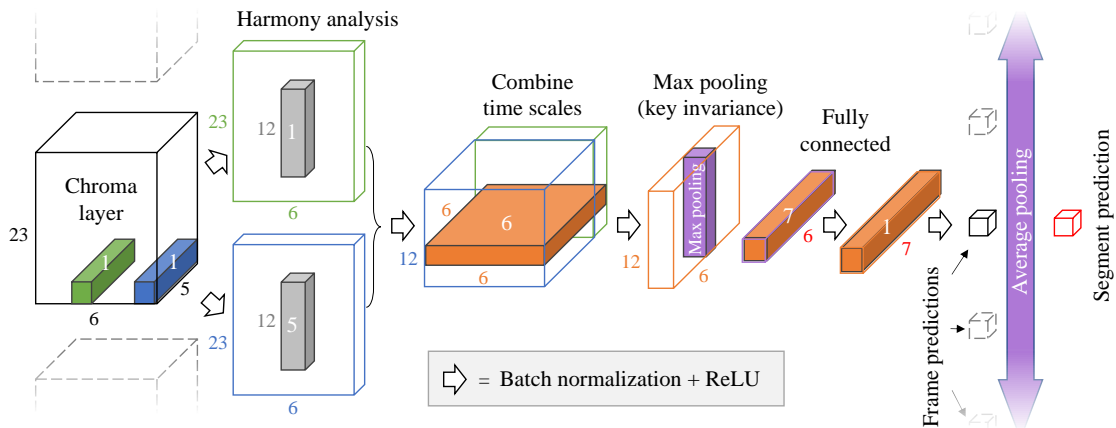

**Figure 2.** The CNN architecture for predicting perceived modality, shown for a single input tensor (time frame). Neighboring time frames are faded and dashed. All dimensions (except singleton) are indicated with black or colored numbers, and the number of filters is indicated with white numbers. White arrows indicate batch normalization followed by ReLUs. In the chroma layer, two filters (green and blue) compute pitch chromas from the input. The two chromas are processed to analyze the relationship between different pitch classes. Various time scales are then combined through six filters (orange), and max pooling applied (purple) to provide invariance with regards to the key class of the music. A fully connected layer is implemented through filters (orange) that span the entire feature space for a specific time frame. Finally, one filter is used to compute a frame prediction, and average pooling is applied to generate a prediction for the entire segment (red).

### 3.2 Training

The system was evaluated with 10-fold cross-validation, (Section 5.2). For each fold, the CNN was trained with the Adam optimizer [21], using the mean-squared-error loss function, an initial learning rate of 0.01, and a drop factor (every epoch) of 0.98. The gradient decay factor was set to 0.9 and the factor for $L_2$ regularization was set to 0.0001. A mini-batch size of 32 (segments) was used, shuffling the training data every epoch. The CNN was trained for 25 epochs. If the computed $R^2$ on the *training split* (the tracks used for training in each of the ten folds) was below 0.83 after 25 epochs, the network was reinitialized and training restarted (to avoid networks stuck in a local minimum). This happened in around 3 % of the cases. We tested the main model without the restarting condition in the ablation study (Section 7.1), which produced a minimal difference. Note that there was no validation stopping, for the same reasons as outlined in [10]: small validation sets are unreliable performance indicators, and maximizing performance for individual networks will not necessarily maximize performance of ensembled (Section 4) networks.

### 4. GLOBAL ESTIMATES AND ENSEMBLING

Two different global estimates were computed for each ME, one estimate from an ensemble of the CNN (ECNN), and one estimate from an ensemble of MLPs (EMLP) refining the output activations from each CNN.

### 4.1 CNN

The CNN modality prediction for each ME was computed as the average of all frame predictions (see Figure 2). This means that the local CNN architecture up until average pooling was applied to each frame at run-time.

We used ensemble learning to improve the accuracy of the CNN predictions (Section 1.4). Ten CNNs were trained for each fold, and the average of their predictions was used.

### 4.2 Additional MLP

In addition, another global estimate was computed for each ME with an EMLP, using the global CNN prediction and input features from the Pitchogram and spectrogram. The intention was to use and examine the effect of various (potentially confounding) features in the audio that the CNN was designed to not model.

Pitch activations in the tuned Pitchogram were averaged across time and summed to a vector, indexing into the vector based on each bins' distance to the closest semitone (0-50 cents). The 6 first discrete cosine transform (DCT III) components of the vector were then extracted as features. These features capture the extent and shape of micro-tuning deviations (PT) across the track (e.g., from vibrato).

We also computed both a vibrato suppressed (VS) and vibrato enhanced (VE) spectral flux (SF), using the max-filtering processing strategy first described in [9], developed to model perceived speed from onset densities in pitched instruments. The processing was applied to the whitened log-frequency signal level spectrogram computed as described in [8]. The VE SF was computed by subtracting the VS SF from the regular half-wave rectified bin-wise SF, thereby retaining energy only in the bins suppressed in the VS SF. For both versions, we computed the mean across time after half-wave rectification, producing an SF vector across frequency. The 6 first DCT components were extracted from these vectors as features. We then used the same log-frequency spectrogram, computed the mean across time, and extracted the first 6 DCT components of the resulting vector (spectral distribution, SD).

The average CNN output activation from all frames (Section 4.1) was also used as an input feature (naturally, this was the most important feature). In summary, the MLP had $1 + 6 \times 4 = 25$ input features, divided into four feature groups (PT, VS, VE and SD) and one CNN prediction.

The MLP was rather small and resembled the MLP developed for predicting performed dynamics in [10]. It had two hidden layers each consisting of 8 neurons. The network was trained for 5 epochs with the Levenberg-Marquadt optimization [30]. Hyperbolic tangent (tanh) units were used as activation functions in all layers except for the last linear output activation. Each input feature was normalized by its minimum and maximum value to the range ±1. Ensemble learning was used, taking the average prediction of 20 MLP models. Since the ECNN consisted of 10 CNNs, and since one EMLP was trained for each CNN, the final prediction in each fold was computed as an average of 10 × 20 = 200 MLP models.

## 5. DATASETS AND EVALUATION PROCEDURE

### 5.1 Datasets

The dataset for the study was assembled from two music audio datasets. The first dataset ($D_1$) consists of 100 audio excerpts of popular music (average length 30 s) that were produced from MIDI [14]. The second dataset ($D_2$) from [4] consists of 110 audio excerpts of film music (average length 15 s). As previously noted [7], the film music dataset contains duplicates. Seven duplicates were found and removed, reducing the size of $D_2$ down to 103 MEs. The MEs are polyphonic and use a wide range of instruments.

The overall modality had previously been rated by two groups of 19 and 21 listeners for the two datasets. Listeners were asked to rate the modality of each excerpt on a quasi-continuous scale between minor (1) and major (10), listening on high-quality loudspeakers. The ratings were averaged across listeners, producing a single ground truth rating of perceived modality for each ME. Reliability was relatively high, with a standardized Cronbach's alpha (CA) [3, 11] of 0.94 and 0.97 for the two datasets.

The datasets were pooled into a single dataset (203 MEs), which was used for training and testing.

### 5.2 Evaluation Procedure

The accuracy of the model was computed with the coefficient of determination, $R^2$, between predictions and ground truth annotations. We used the square of Pearson's correlation coefficient (including an intercept).

The models were evaluated with 10-fold cross-validation, using a stratified sampling so that each training set contained about the same number of MEs from $D_1$ and $D_2$. To improve the reliability of the results, the *complete experiment* was repeated ten times, re-partitioning the validation split each time.[2] To get 95 % confidence intervals (CIs), ten results ($R^2$s) were sampled with replacement and the mean computed, repeating the procedure $10^6$ times. The resulting distribution of mean $R^2$s could then be used for extracting CIs; it indicates the reliability of the test results based on its variation over test runs.

## 6. RESULTS

### 6.1 Main Results

The final result for the ECNN and the ECNN in combination with the global EMLP is presented in Table 1.

---

[2] Running the main 10-fold cross-validation experiment ten times took about 5.5 days, training the ECNN with a GeForce GTX 1080 GPU (5 days; 10 × 10 × 10 = 1 000 CNNs) and the EMLP using 5 parallel i7-6700K CPU threads (0.5 days; 10 × 20 × 10 × 10 = 20 000 MLPs).

| Model | $R^2$ | 95 % CI | $D_1$ | $D_2$ |
|---|---|---|---|---|
| ECNN | 0.672 | 0.665-0.679 | 0.645 | 0.710 |
| ECNN+EMLP | 0.716 | 0.710-0.722 | 0.710 | 0.745 |

**Table 1.** Squared correlation ($R^2$) between the ground truth ratings of perceived modality in music audio and the predictions of the two proposed models; also measured individually within the two datasets ($D_1$ and $D_2$).

The full system reached an $R^2$ of 0.716 for the predictions of perceived modality in the two datasets (corresponding to a correlation, $r$, of 0.846). The predictions from the CNN ensemble without the subsequent EMLP, minimizing contributions from confounding factors of variation, were almost as accurate, with an $R^2$ of 0.672.

As seen in Table 1, the second dataset ($D_2$) consisting of film clips was easier to predict than the first one ($D_1$). This difference is in line with results of the previous study on the same datasets that reported an $R^2$ of 0.43 (full model, 0.38) for $D_1$ and 0.47 (full model, 0.53) for $D_2$. The higher CA (Section 5.1) for this dataset indicates that listeners also had stronger agreement when rating it. Figure 3 shows predictions in relation to ground truth annotations.

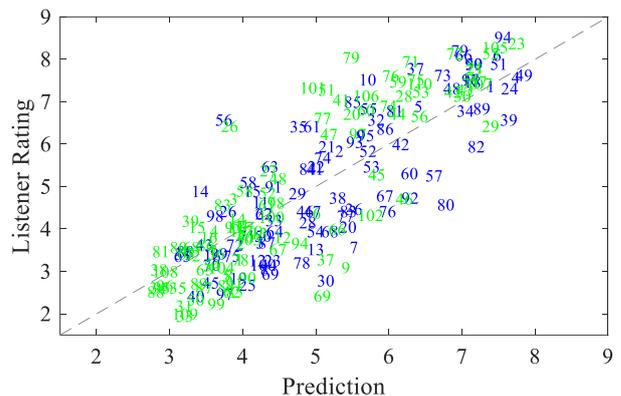

**Figure 3.** Predicted modality (x-axis) in relation to rated modality (y-axis) for datasets $D_1$ (blue) and $D_2$ (green) in one of the test runs ($R^2 = 0.706$) for the ECNN+EMLP. The dashed grey line indicates perfect prediction. Numbers indicate the index of each ME for future comparisons.

### 6.2 Comparison with Previous Systems and Humans

Figure 4 provides context regarding the prediction accuracy. The proposed system (blue bars) clearly outperforms previous systems (Section 1.2, white bars). Note that [1] was tested on a different dataset, so differences in predictive performance should be interpreted with caution. Human performance (circles) was computed from the listeners of the original listening test using a similar strategy as proposed in [10]. The performance of $n$ listeners was derived by sampling (with replacement) $n$ listeners and computing the $R^2$ between their mean rating and the mean rating from the non-sampled listeners. The procedure was repeated $10^5$ times, using the $10^5$ results to compute a mean and 95 % CIs. For $n = 1$, sampling is not applicable, and the 95 % CIs were defined as the listener with the second lowest and second highest $R^2$.

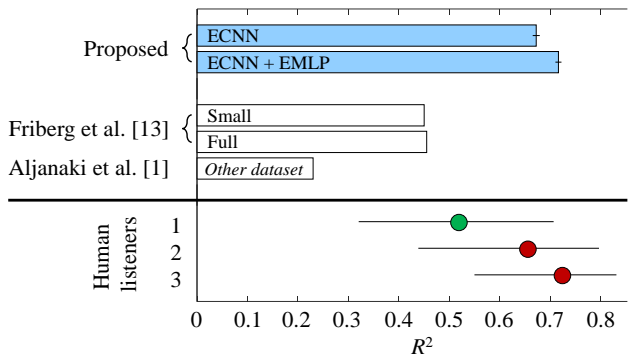

**Figure 4.** Modality estimation results ($R^2$) of the proposed system (blue), previous systems (white), individual human listeners (green circle), and ensembled human listeners (red circles). Black lines indicate 95 % CIs.

## 7. ABLATION STUDY

### 7.1 CNN

Important stages of the CNN architecture were examined by training the ECNN with various alterations of the CNNs. The different stages/properties tested were:

**Time scales:** *None-2431* – All different time scales (Section 2.4) were tested separately.

**Key class invariance pooling (Pool):** *Avg* – A CNN using average pooling instead of max pooling. *Conv* – A fully convolutional architecture that instead reduced the pitch dimension through four layers with two filters of height {5 4 3 3}, followed by a single filter of height 1.

**Input:** *Mag* – Using a magnitude log-frequency spectrogram as input, computed as described in [8] (pre-filtering). *dB* – Using the whitened log-frequency signal level spectrogram from [8]. For both versions, the spectrum covered the same range of 71 semitones, using overlapping triangular filters to reduce the frequency resolution.

**Pitch chroma (PC):** *Mean* – The mean of the five octaves was computed directly and passed as input, using 6 harmony analysis filters in the first layer

Results relative to the main CNN model are shown in Figure 5 and conclusions of the experiment provided in Section 8. The same validation split was used for all tests to increase consistency. This split was also used for the main model for computing a performance reference.

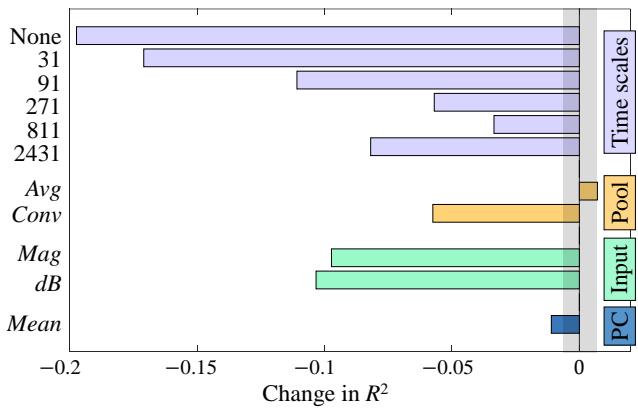

**Figure 5.** The variation in $R^2$ for different CNN settings in relation to the main CNN model. The 95 % CIs ($\pm 0.007$) from the main experiment are indicated by the grey area.[3]

---

[3] Note that these CIs define the 95 % range within which the mean of *ten* complete 10-fold cross-validation runs varies. The ECNN ablation study used *one* complete 10-fold cross-validation run per architecture.

### 7.2 MLP

We also tested various combinations of EMLP input features.[4] The results shown in Figure 6 indicate that a combination of SF features with vibrato suppression (VS) and vibrato enhancement (VE) was important.

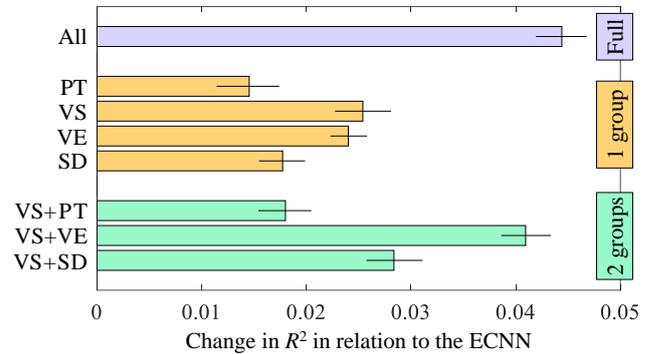

**Figure 6.** The change in $R^2$ in relation to the ECNN results, when using various EMLP feature groups (Section 4.2). Black lines indicate 95 % CIs for the relative improvement over the ECNN in each test run.

## 8. CONCLUSIONS

A convolutional neural network for predicting perceived modality in music was implemented. Its predictive performance was well above that of previous systems as well as the average human listener, performing better than around 95 % of the human annotators. It requires the combined ratings of 3 listeners to reach the same predictive performance as the model. The CNN used pitch activations from a pitch tracking system as input; the ablation study showed that this input representation improves performance substantially in relation to spectral input (*Mag* and *dB*). The methodology of max pooling across key classes to provide invariance seems beneficial since it improved performance in relation to a fully convolutional model. However, average pooling, in which the network instead has to rely on earlier ReLUs to discard irrelevant activations in certain key classes seems to be an equally attractive, or even better, option for achieving key class invariance.

The CNN was restricted from using filters operating across time, to reduce the influence of irrelevant confounding factors of variations, such as accentuation and spectral fluctuations. Instead, the CNN received input filtered to account for different time scales. The ablation study indicates that a time scale of around 4-5 seconds is the most relevant and that instantaneous time scales, only using harmonic information from concurrent tones, significantly reduces performance. Performance only dropped slightly when the pitch chroma layer of the CNN was discarded and the mean (across octaves) pitch chroma instead used as input (*Mean* PC, Figure 5). The small size of the dataset likely reduces the importance of this CNN layer; tracking interactions between pitches in different registers requires more learnable parameters, which requires more input data for generalization.

We hope that the results can inspire further development of CNN architectures accounting for musical invariances, including, and beyond, key class and pitch class.

---

[4] All feature groups were tested with the *same* global CNN activation as an additional feature, and evaluated across the full ten experimental runs.